\documentclass[a4paper]{jpconf}
\usepackage{graphicx}
\begin{document}
\title{Quantum entanglement in the initial and final state of relativistic heavy ion collisions}

\author{Rene Bellwied}

\address{Physics Department, University of Houston, 617 SR1 Building, Houston, TX 77204, USA}

\ead{bellwied@uh.edu}

\begin{abstract}
The possibility of quantum entanglement leading to a seemingly thermal distribution of the initial partonic state that maps to the final hadronic state in the evolution of the deconfined phase generated in relativistic heavy ion collisions is discussed in the context of early thermalization and final state particle distributions obtained in experiments at RHIC and the LHC.
\end{abstract}

\section{Introduction}

The creation of the deconfined phase of QCD, the Quark-Gluon Plasma (QGP), in heavy-ion collision experiments at RHIC and the LHC has generated increasing theoretical activity, aimed at understanding the quantum properties of matter under extreme conditions. It is believed that the deconfined system initially is in a gluon condensed state that evolves to a quark-gluon phase before transitioning to a confined hadronic phase. At high temperatures and low chemical potential, $\mu_B$, this transition is an analytic crossover \cite{Aoki:2006we}, which is followed by a hadronic phase that is bracketed by the chemical freeze-out on the high temperature side and the kinetic freeze-out on the low temperature side, assuming thermal equilibrium has been reached at the time of hadronization. The notion of common freeze-out surfaces for all produced particles requires the system to be in, at least, local equilibrium at the onset of the hadronic phase, but furthermore the commonly accepted hydrodynamical description of the preceding partonic phase requires the quantum system to be thermalized at very early times ($<$ 1 fm/c). Since such early thermalization is likely impossible to reach through interactions between the parton constituents, several interesting ideas have been proposed, ranging from plasma instabilities leading to thermalization to Hawking-Unruh radiation giving the impression of a thermalized system \cite{Strickland:2007fm, Castorina:2007eb}. Here we investigate the idea that seemingly local thermalization in the initial state could be attributed to quantum entanglement and we pose the question whether the entanglement could survive the subsequent system evolution and ultimately lead to the final state particle multiplicity distributions, in particular for composite light nuclei and multi-quark states, which were measured at RHIC and LHC. 

The underlying theory is based on recent measurements in an isolated quantum many body system made of ultracold atoms \cite{kaufman}. Here it was found that a local disturbance can lead to apparent thermalization through quantum entanglement for an isolated subset of the system. A series of recent theory papers aimed at applying the same theory to the emission spectrum obtained in, either deep inleastic scattering or from an expanding quantum string \cite{Kharzeev:2017qzs,Berges:2017hne}, in order to explain the apparent thermal emission obtained in elementary relativistic e$^{+}$e$^{-}$ and pp collisions \cite{Baker:2017wtt}.
The goal of this manuscript  is to use a similar argument in order to obtain a microscopic understanding of the high yields of composite objects, such as light or hypernuclei, in heavy ion collisions at RHIC and LHC. The novel idea here is that inital quantum entanglement in the QCD system does not only lead to thermal particle spectra, but it also defines the partonic composition and the abundance of each final state hadron, up to the most composite quark structure. The particles in question, namely light or hyper-nuclei, feature binding energies in the hundreds of keV range, but their yield is in agreement with calculations based on a thermally equilibrated system of a temperature around the QGP transition temperature, i.e. around 150-160 MeV \cite{Andronic:2010qu}. These original results at RHIC energies were recently confirmed through detailed measurements with ALICE at the LHC \cite{Adam:2015vda}. The production mechanism is not clear since  purely hadronic coalescence models, which only require a wave function overlap between baryons near the kinetic freeze-out, i.e. at much lower temperatures when the hadronic interactions cease, yield similar cross sections for light nuclei than the thermal models. In a thermal many body system near the QCD transition it is  rather inconceivable that this loosely bound objects would survive the surrounding heat bath. On a macroscopic level it was found, though, that the thermodynamic entropy per baryon in a fireball, is  fixed when the system is in its most compressed state \cite{Siemens:1979dz,Csernai:1986qf}, although no  microscopic explanation for this phenomenon exists. Here we try to argue that the thermodynamic entropy of the final state can be related to the entanglement entropy in the initial state and thus quantum entanglement drives the final state particle multiplicities up to the heaviest measured particles, namely the light nuclei at RHIC and the LHC, even in heavy ion collisions.

\section{The initial state}

The idea of applying quantum entanglement to the evolution of a parton collision inside a proton-proton system seems natural since any parton in the color neutral proton should be entangled transversely with the other valence quarks through a density matrix that includes mixed states. In particular, the sum of all hadronic final states have to contain mixed states since the entanglement of the valence quarks leads to several components of the proton's wave function \cite{Kharzeev:2017qzs}. This type of quantum entanglement can be viewed as transverse entanglement and should be complemented with a longitudinal component that builds up when determining the particle production from a dynamically evolving QCD string produced in the collision. For example in PYTHIA, the different regions in a string are entangled in rapidity space and any given region is described by a mixed state reduced density matrix \cite{Berges:2017hne}. Overall the combined longitudinal and transverse quantum entanglement makes the entanglement entropy volume dependent and thus an extensive quantity. The coherent state vacuum at the onset of the collision can therefore be described by entangled pairs of quasi-particles with opposite wave numbers. Since at these times in the evolution the system is strongly coupled we can apply a conformal field theory, which yields a relationship between entropy and time-dependent temperature that was first deduced by Calabrese and Cardy \cite{Calabrese:2004eu}, The following equations should be valid locally, where the local extent is determined by the volume over which the entanglement holds, i.e.

\begin{equation}
L = \tau \Delta\eta
\end{equation}
in this case the time-dependent temperature becomes
\begin{equation}
T =\frac{1}{2\pi\tau}
\end{equation}
where the entropy is defined as
\begin{equation}
S(\tau, \Delta\eta) = \frac{c}{3} ln (\frac{2\tau}{\epsilon} sinh(\Delta\eta/2)) + const.
\end{equation}
see also \cite{Berges:2017hne}. As can be seen from the equations above, the temperature falls when the time $\tau$, and thus the entanglement volume, increases. 

 \section{Dynamical evolution from the initial state to the phase transition}

The question is whether this type of quantum entanglement could determine the thermal-like yields and spectra of the final state hadrons. In the calculations the quantum entanglement is spatially motivated, meaning the density matrices are given by the parton distributions which relate simply to the entanglement entropy. If the density is high, the system consists of many states all with equal probability and thus maximally entangled. This is easiest visualized for low-x where the system can be considered a pure maximally dense gluon state. The picture becomes more complicated if we allow quarks to contribute to the number of states, and thus one first propagates a pure glue system through the partonic phase of the QGP. This can be done through a quenched lattice QCD calculation. In the end the quark degrees of freedom need to contribute to the number of entangled states, also in order to push the volume of the entangled system and the hadronization temperature to realistic values. For a maximally entangled state the final volume reflects the fireball at hadronization, and thus it is no surprise that the resulting temperature is close to the Hagedorn temperature. If one assume parton-hadron duality then this system could potentially freeze-out into hadrons and the entanglement entropy would simply translate into a thermodynamic entropy obtained from the Boltzmann distribution of the final state. This step requires, though, that the entanglement during the dynamic evolution of the partonic system is not destroyed, in other words the system does not decohere through partonic interactions. Any hadronic interactions after freeze-out do not impact the initial particle yield since the emission from the quantum entangled system determines the confined state. If this principle goes as far as preserving a multi-quark bag which forms composite objects such as light nuclei at freeze-out, then quantum entanglement can serve as an explanation for the seemingly thermal yields of light nuclei that have binding energies that are several orders of magnitude lower than the thermalization temperature at freeze-out.

\section{Experimental evidence in the final state}

Historically parton-hadron duality was used to explain the perturbative part of the spectrum  in elementary collisions \cite{Shifman:2000jv}, but the apparent similarity between the gluon densities required to explain the nuclear suppression factors measured in relativistic heavy-ion collisions at RHIC and the number of charged hadrons measured in the detector can also lead to the application of the model in larger multi-body systems \cite{Dokshitzer:1991eq}. In this picture the gluon saturation scale will determine the final state hadron multiplicities. and lead to a geometrical scaling as a function of an energy dependent saturation scale \cite{Iancu:2003xm,Kharzeev:2001yq}. Although this scaling intially seemed to be broken in heavy-ion collisions at LHC energies, new studies  point out that, when taking into account the effect of DGLAP evolution on the scale, an agreement with experimental data can be reached which preserves the idea of LHPD \cite{Lappi:2011gu}.

Any LPHD scenario neglects all collective effects in AA collisions, and therefore should be viewed as the opposite of a model describing a system that expands in the transverse and longitudinal direction according to ideal hydrodynamical equations of motion. Interestingly, though, in the ideal hydrodynamical case and in LHPD the entropy, and thus the multiplicity of the relevant degrees of freedom stays constant during the evolution. This leads in both cases to the fact that the final particle  multiplicity is effectively equal to the initial gluonic multiplicity. Although appealing as a concept, for any hydrodynamical model there is still no quantitative theoretical understanding of the thermalization process. Here the quantum entanglement argument, which leads to an entanglement entropy that maps onto the thermodynamic entropy, offers an intuitive picture.

\subsection{Violation of KNO scaling in elementary collisions}

Experimental evidence for the potential equivalence between entanglement entropy of the initial system and the von Neumann (thermodynamic) entropy of the final system can be verified through  rather well established experimental effects in particle distributions. Without explicitly determining the gluon density one can measure a breakdown of the scaling relation between the final hadron multiplicity and the number of scatterings for each parton (KNO scaling) \cite{Koba:1972ng}. If KNO scaling holds, then the scaling of particle production would be independent of the process. Recent LHC results have confirmed KNO scaling even at the highest energies as long as the pseudo-rapidity window is small. A larger coverage, though, leads to the breakdown of the process independent scaling, which is well documented by plotting the energy dependence of the higher cumulants of the multiplicity distribution  \cite{Khachatryan:2010nk}. There are large multiplicity, and thus cumulant, fluctuations in the soft sector which lead to non-Poissonian cumulants.  Any entanglement in the soft gluon sector leads thus to a violation of the KNO scaling, and the predictions based on gluon saturation and enhanced soft multi-parton scattering are in good agreement with the measured cumulants. 

\subsection{Thermal emission of composite objects at the QCD phase transition}

The mapping between the inital gluon density and the final hadron density is also realized  in heavy ion systems in the context of the aforementioned semi-classical gluon saturation model to describe the initial state, i.e. the so-called Color Glass Condensate (CGC) \cite{Iancu:2003xm}.  It is important to realize that the scaling violations determined in elementary pp collisions can also be applied to understand parton-hadron duality in the heavy ion system, although also in the heavy ion system the duality needs to be extended from the perturbative regime to the highly non-perturbative part of the spectrum, i.e. soft gluon contributions will dominate the final particle spectrum. A more detailed analysis of the higher moments of the multiplicity distribution as a function of system size and collision energy can certainly shed light on the link between saturation level and final charged particle multiplicity.

The advantage that a heavy ion particle distribution has to offer is that the abundance of rare particle species enables the application of  a thermal fit over many more orders of magnitude.The rather tenuous link to thermal behavior in elementary collisions \cite{Becattini:2010sk} thus gets replaced with solid fits based on statistical equilibrium models \cite{Stachel:2013zma}. If the yield of rare light nuclei and hypernuclei indeed continues to follow the trend of reflecting a particular freeze-out temperature, which is identical to the chemical freeze-out of all basic particle species, as was shown recently by the ALICE collaboration \cite{Adam:2015vda}, then the measured temperature could be connected to the entanglement entropy as shown in Eqs.1-3.The 'temperature' then reflects a degree of quantum entanglement, which is described by the entanglement entropy specific for a particular time and volume  near the hadronization transition. An alternative explicit link between the entanglement 'temperature' and the thermal temperature measured from the spectra of particles was derived in \cite{Feal:2018ptp}, where the gluonic system is frozen in early in the evolution through a rapid quench that generates a highly excited multi-particle state.The thermal spectrum  originates from the event horizon formed by the acceleration of the color field. Any mapping between the calculated temperature and the temperature obtained from the final hadron spectra then relies on the clustering of color sources during hadronization. This theory tries to link the kinetic freeze-out temperatures to the entanglement 'temperature' whereas here we link the particle multiplicities at the chemical freeze-out temperature, i.e. the thermodynamic entropy per baryon, to the entanglement entropy.

The evolving picture is that the dynamical evolution of the initial saturation state through the strong coupling regime, during the QCD crossover, is likely not changing the entropy of the system very much, in which case a one to one correspondence between the initial entanglement entropy and the final thermodynamic entropy could be postulated. This gives a microscopic explanation to the early findings by Siemens and Kapusta \cite{Siemens:1979dz} that showed that, at lower energies near the nuclear liquid-gas phase transition, the entropy per baryon of the final system is apparently fixed in the compressed and hot state of the produced fireball, which led later on to the 'snowball in hell' interpretation by Braun-Munzinger and Stachel \cite{BraunMunzinger:1994iq}.

Unfortunately the interpretation of the measured light nuclei abundances at the LHC is not unique. The thermal model  emission yields are very close to coalescence predictions. Coalescence is a process that solely requires sufficient wave function overlap among hadrons in the final state, and thus no equilibrated system,  in order to yield composite objects, whereas in the thermal model the production yield depends on the final chemical freeze-out parameters (T and $\mu_{b}$).  The entanglement scenario would be closer to the thermal production in that it yields a common parameter, T, for all particle species, although it gives this parameter a different meaning. Here the apparent temperature from a thermal equilibrium fit will  only reflect the initial maximum level of entanglement, which would stay constant as a function of collision energy as soon as the saturation scale is reached. Therefore it is not surprising that the chemical freeze-out temperature follows the Hagedorn temperature and becomes energy independent at RHIC and LHC energies. The formation of light nuclei will then reflect the longitudinal and transverse entanglement in the initial QCD phase but the yields can still be predicted by a model assuming thermalization at a fixed temperature.In this context it is experimentally relevant that recent calculations, based on a common emission temperature, show that multi-strange multi-quark objects or hypernuclei are within range of the high statistics runs at the LHC scheduled from 2021 on \cite{Dainese:2016dea}.

\section{Outlook}

The model of quantum entanglement of the initial state produced in a relativistic heavy ion collision might give rise to complex particle production in the final hadronic state. The entanglement entropy dominated by the soft gluon sector in a gluon saturated system could be matched one to one to a thermodynamic entropy of the final state hadron system. Since the entropy per baryon would then be driven by the entropy per gluon any complex baryonic structure could be traced back to an initial QCD string that carries its own quantum entanglement properties based on the mixed state density matrix of the initial state. 

One of the more far-reaching consequences of such an approach to particle formation in the QCD crossover is the fact that nuclei production up to a certain complexity of the final state is determined by the initial state saturation scale. At which point the initial state entanglement is exhausted needs to be calculated in a more detailed approach.

\bigskip

\section{Acknowledgments}

This work is supported by the U.S. Department of Energy under Contract No. DEFG02-07ER41521


\section*{References}

\begin{thebibliography}{999}

\bibitem{Aoki:2006we}
  Y.~Aoki, G.~Endrodi, Z.~Fodor, S.~D.~Katz and K.~K.~Szabo,
  Nature {\bf 443}, 675 (2006)


\bibitem{Strickland:2007fm} 
  M.~Strickland,
  J.\ Phys.\ G {\bf 34}, S429 (2007)


\bibitem{Castorina:2007eb} 
  P.~Castorina, D.~Kharzeev and H.~Satz,
  Eur.\ Phys.\ J.\ C {\bf 52}, 187 (2007)


\bibitem{kaufman}
A.M. Kaufman,M.E. Tai, A.Lukin, M. Rispoli, R.Schittke, P.M. Presiss, M. Greiner
Science 353, 794 (2016) 

\bibitem{Berges:2017hne} 
  J.~Berges, S.~Floerchinger and R.~Venugopalan,
  JHEP {\bf 1804}, 145 (2018)


\bibitem{Kharzeev:2017qzs} 
  D.~E.~Kharzeev and E.~M.~Levin,
  Phys.\ Rev.\ D {\bf 95}, no. 11, 114008 (2017)
 

\bibitem{Baker:2017wtt} 
  O.~K.~Baker and D.~E.~Kharzeev,
  arXiv:1712.04558 [hep-ph].

\bibitem{Andronic:2010qu} 
  A.~Andronic, P.~Braun-Munzinger, J.~Stachel and H.~Stocker,
  Phys.\ Lett.\ B {\bf 697}, 203 (2011)

\bibitem{Adam:2015vda} 
  J.~Adam {\it et al.} [ALICE Collaboration],
  Phys.\ Rev.\ C {\bf 93}, no. 2, 024917 (2016)
 

\bibitem{Siemens:1979dz} 
  P.~J.~Siemens and J.~I.~Kapusta,
  Phys.\ Rev.\ Lett.\  {\bf 43}, 1486 (1979).


\bibitem{Csernai:1986qf} 
  L.~P.~Csernai and J.~I.~Kapusta,
  Phys.\ Rept.\  {\bf 131}, 223 (1986).

\bibitem{Calabrese:2004eu} 
  P.~Calabrese and J.~L.~Cardy,
  J.\ Stat.\ Mech.\  {\bf 0406}, P06002 (2004)

\bibitem{Shifman:2000jv} 
  M.~A.~Shifman,
  hep-ph/0009131.

\bibitem{Dokshitzer:1991eq} 
  Y.~L.~Dokshitzer, V.~A.~Khoze and S.~I.~Troian,
  J.\ Phys.\ G {\bf 17}, 1585 (1991).

\bibitem{Iancu:2003xm} 
  E.~Iancu and R.~Venugopalan,
  In *Hwa, R.C. (ed.) et al.: Quark gluon plasma* 249-3363
  [hep-ph/0303204]

\bibitem{Kharzeev:2001yq} 
  D.~Kharzeev, E.~Levin and M.~Nardi,
  Phys.\ Rev.\ C {\bf 71}, 054903 (2005)


\bibitem{Lappi:2011gu} 
  T.~Lappi,
  Eur.\ Phys.\ J.\ C {\bf 71}, 1699 (2011)

\bibitem{Koba:1972ng} 
  Z.~Koba, H.~B.~Nielsen and P.~Olesen,
  Nucl.\ Phys.\ B {\bf 40}, 317 (1972)

\bibitem{Khachatryan:2010nk} 
  V.~Khachatryan {\it et al.} [CMS Collaboration],
  JHEP {\bf 1101}, 079 (2011)

\bibitem{Becattini:2010sk} 
  F.~Becattini, P.~Castorina, A.~Milov and H.~Satz,
  Eur.\ Phys.\ J.\ C {\bf 66}, 377 (2010)

\bibitem{Stachel:2013zma} 
  J.~Stachel, A.~Andronic, P.~Braun-Munzinger and K.~Redlich,
  J.\ Phys.\ Conf.\ Ser.\  {\bf 509}, 012019 (2014)

\bibitem{Feal:2018ptp} 
  X.~Feal, C.~Pajares and R.~A.~Vazquez,
  arXiv:1805.12444 [hep-ph].

\bibitem{BraunMunzinger:1994iq} 
  P.~Braun-Munzinger and J.~Stachel,
  J.\ Phys.\ G {\bf 21}, L17 (1995)

\bibitem{Dainese:2016dea} 
  A.~Dainese {\it et al.},
  Frascati Phys.\ Ser.\  {\bf 62} (2016)
  [arXiv:1602.04120 [nucl-ex]].


\end{thebibliography}
\end{document}